\documentclass[twocolumn,showpacs,preprintnumbers,amsmath,amssymb]{revtex4}

\usepackage{graphicx}

\begin{document}

\title{Observation of a strongly nested Fermi surface in the shape-memory
alloy Ni$_{0.62}$Al$_{0.38}$}

\author{S.~B. Dugdale, R.~J. Watts, J. Laverock, Zs. Major,
M.~A. Alam}
\affiliation{H.~H.~Wills Physics Laboratory, University of Bristol, Tyndall
Avenue, Bristol BS8 1TL, United Kingdom}

\author{M. Samsel-Czeka{\l}a, G. Kontrym-Sznajd}
\affiliation{W. Trzebiatowski Institute of Low Temperature and Structure Research,
Polish Academy of Sciences, P.O. Box 1410, 50-950 Wroc{\l}aw 2, Poland}

\author{Y. Sakurai, M. Itou}
\affiliation{Japan Synchrotron Radiation Research Institute, SPring-8, 1-1-1 Kouto,
Mikazuki, Sayo, Hyogo 679-5198, Japan}

\author{D. Fort}
\affiliation{School of Metallurgy and Materials, University of Birmingham,
Birmingham B15 2TT, United Kingdom}

\begin{abstract}

The Fermi surface topology of the shape-memory alloy Ni$_{0.62}$Al$_{0.38}$ has 
been determined using Compton scattering. A large area of this Fermi surface 
can be made to nest with other areas by translation through a vector
of $\approx 0.18~[1,1,0](2\pi/a)$, which correponds to the wavevector associated
with martensitic precursor phenomena such as phonon softening and
diffuse streaking in electron diffraction patterns. This observation
is compelling evidence
that these phenomena are driven by the enhanced electron-lattice coupling
due to the Fermi surface nesting.

\end{abstract}


\maketitle

Smart alloys which exhibit shape-memory and super-elastic phenomena have
been deployed in a wide variety of applications ranging from actuators in
aircraft wings to surgical instruments.
However, an atomic-scale understanding of the origin of the martensitic
transformation (MT), the structural transformation at the heart of these
phemomana, is still lacking. It has been hypothesised that lattice
vibrations are the key, an idea supported
by first-principles calculations indicating that strong coupling
of certain phonons to the
electrons (phonon softening), due to particular features in the Fermi surface,
plays a crucial role
\cite{zhao:92,naumov:97,isaev:04,huang:03,huang:04,bungaro:03}.
Owing principally
to the compositional disorder inherent to many of these alloys,
a Fermi surface determination in these materials is experimentally
challenging, with traditional quantum oscillatory techniques suffering due to
their reliance on a long electronic mean free path.
Recent de Haas-van Alphen experiments in the austenitic phase of the
low-temperature shape-memory alloy AuZn have reported an orbit whose
cross-sectional area is in excellent agreement with first principles
calculations which predict the presence of a strongly nested sheet of
Fermi surface \cite{mcdonald:05}.
In this Letter we provide experimental evidence in support of the
intimate relationship between the phonon softening and 
the Fermi surface through a Compton scattering study of the 
shape memory alloy Ni$_{0.62}$Al$_{0.38}$.

It is more than a century since Martens' exploration of the microstructure
of steels gave the first insight into the origin of their macroscopic
properties, and subsequently led to his name being associated with a
more general solid-state phase transformation \cite{nishiyama:78}.
While metallurgists have a precise, albeit phenomenological, definition
of a MT, and physicists use the term more loosely to define many first-order
transitions with acoustic anomalies, both disciplines generally agree that
they are purely structural, diffusionless
(meaning no atom-by-atom jumping within the unit cell),
displacive (meaning a coordinated displacement of the atoms over distances
much less than the atomic spacing) solid-solid phase transitions
\cite{shapiro:91}. However,
MTs have continued to fascinate generations of scientists.
Ni$_{x}$Al$_{1-x}$ alloys are well known to exhibit a MT for $x \sim 0.62$
\cite{zhao:92},
and represent a prototypical compound for the transformation which
is intermediate between weakly (almost second order) and strongly first order.

Early attempts to understand these transformations suggested that a
phonon would become unstable at a particular temperature, at which point the 
lattice would displace spontaneously to a finite amplitude
\cite{cochran:60,anderson:60}. This `soft-mode' theory, however, is
found to be lacking in several key aspects
\cite{krumhansl:89} which include (i) the observation that the phonon
frequencies only soften slightly, and do not indicate harmonic instability,
(ii) the prediction of a second-order transition
whereas in practice there is a finite, albeit small discontinuity in the
microscopic order parameter (the static amplitude of the frozen-in soft
mode), (iii) the empirical absence of critical fluctutations,
and finally (iv) the
presence of strong precursors of the new phase, or a distorted form of it,
which persist well above the transition temperatures; examples of these
precursors in the case of Ni$_{x}$Al$_{1-x}$ include
diffuse streaking in electron diffraction patterns and the
corresponding `tweed' strain-contrast patterns in transmission electron
microscope images, and anomalous softening of the transverse
acoustic (TA$_{2}$) phonon branch along $[\xi \xi 0]$ at
temperatures well above the MT \cite{shapiro:86,shapiro:91}.
These premartensitic phenomena have been
associated with local deviations from the perfect cubic structure
at temperatures well above the MT, the regions being formed
as a result of the coupling of defect-induced
strain fields and anomalous phonon softening.

It is well-known
that when parallel pieces of Fermi surface exist in a metal, 
there will be a strong electronic response at the wavevector which
translates, or nests, one parallel piece onto the other. This role
of the Fermi surface in influencing the electron-phonon coupling,
was extensively investigated during the last two decades (see, for example,
\cite{bruinsma:82,zhao:89,zhao:93}), where premartensitic phenomona 
\cite{shapiro:86,shapiro:91} were explicable in terms of Fermi surface
nesting.

While anharmonic effects are thought
to be responsible for the MT \cite{ye:87}, it is Kohn anomalies
\cite{kohn:59} driven by nestable regions on the Fermi surface
impacting on the electronic screening (and hence on the electron-phonon
coupling), that were initially suggested as the origin of the premartensitic
phenomena \cite{zhao:92}. Given the disordered nature of the Ni$_{x}$Al$_{1-x}$
alloys, the degree to which the Fermi surface remains a well-defined,
sharp entity is an important consideration. Stocks {\it et al.}
\cite{stocks:92} tackled this issue by calculating the effect of disorder
within the framework of the
coherent potential approximation (CPA). They were able to
show that although there is significant smearing of the Fermi surface
in many areas of the Brillouin zone, the region of Fermi surface 
first identified by Zhao and Harmon \cite{zhao:92} as exhibiting nesting
(surprisingly, perhaps) remains rather sharp.

\begin{figure}[b]
\includegraphics[width=0.80\linewidth,clip]{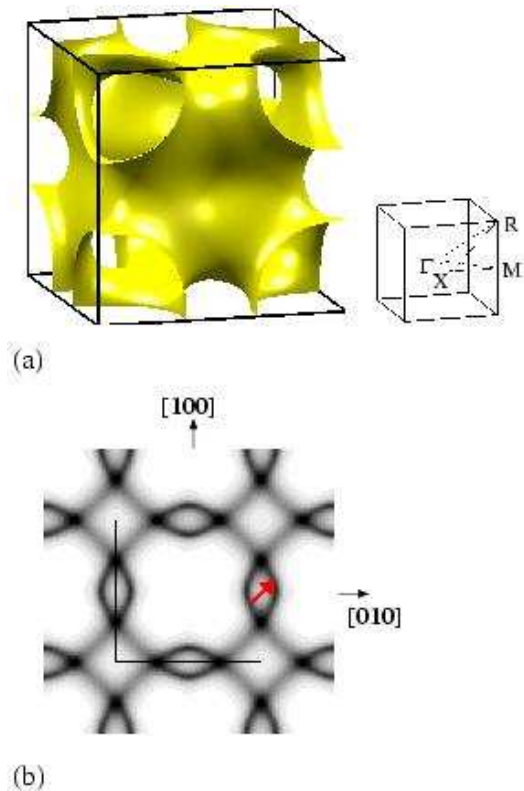}
\caption{\label{kkr} (color online) (a) The Fermi surface of
Ni$_{0.68}$Al$_{0.32}$,
determined via the KKR calculation (shown alongside are the location
of some relevant symmetry points within the Brillouin zone), and
(b) the intensity of the Bloch spectral function,
where dark shades represent high intensity,
of the sheet of Fermi surface proposed to
accommodate the nesting in Ni$_{0.68}$Al$_{0.32}$, shown on the
$k_{z}=0.48~(\pi/a)$ plane. The Fermi surface can be identified as
the strong oval features located on the edges of the Brillouin
zone (the fainter lines being spectral weight from a band close to the
Fermi energy). The arrow indicates the proposed
nesting vector (along the [110] direction) and the edge of the first
Brillouin zone is marked. Note that this
sheet remains relatively sharp throughout the Brillouin zone, despite the
disorder inherent to the system.}
\end{figure}

The key questions addressed here, however,
are whether the actual Fermi surface of Ni$_{0.62}$Al$_{0.38}$ resembles
that predicted by {\it ab initio} calculation \cite{zhao:92,stocks:92},
and whether any nesting vector matches that of the strongly softened phonons. 
A previous attempt (also using Compton scattering) to establish a
connection between the Fermi surface topology and any shape-memory
behavior \cite{shiotani:04} was unsuccessful, principally due to the
complexity of the Fermi surface of their Ti$_{48.5}$Ni$_{51.5}$ alloy.
Here, we present our measurement of the Fermi surface of
Ni$_{0.62}$Al$_{0.38}$ using the technique of Compton scattering 
\cite{cooper:85}.

In order to make a direct comparison with the Fermi surface
predicted by band theory,
we performed {\it ab initio} electronic structure calculations
to reproduce the earlier work of Stocks {\it et al.} \cite{stocks:92}
for the disordered Ni$_{0.62}$Al$_{0.38}$ alloy.
We employed the fully relativistic
Koringa-Kohn-Rostoker (KKR) method within the atomic sphere
approximation,
and the disorder was taken into account by the coherent potential
approximation \cite{Munich}. The lattice constant was taken to be 2.82~\AA~
and convergence was achieved at 816
k-points within the irreducible Brillouin zone. The Fermi surface was
identified by the locus of the peaks in the Bloch spectral function,
$A^{B}({\mathbf k},\epsilon)$. The sheet of (nestable) 
Fermi surface identified by Zhao and Harmon \cite{zhao:92} is presented in
Fig.\ \ref{kkr}, alongside the three dimensional Fermi surface.

Our single crystal sample was cut by spark erosion from a single grain
of a large ingot of Ni$_{0.62}$Al$_{0.38}$ grown using the
Bridgman method. It was subsequently annealed under argon at
temperatures between 1230K and 1270K for a total of 30 hours, followed by a
vacuum degas at 900K.

A total of twenty-four Compton profiles along different crystallographic
directions were measured at room temperature on the high-resolution
Compton spectrometer of beamline BL08W at the SPring-8 synchrotron
\cite{hiraoka:01,sakurai:04}. This spectrometer is a
Cauchois-type spectrometer, consisting of a Cauchois-type crystal analyser
and a position-sensitive detector, with a resolution FWHM at the Compton peak
of 0.155~a.u (1~a.u.\ of momentum~$=$~1.99~$\times$10$^{-24}$~kg~m~s$^{-1}$)
\cite{hiraoka:01,sakurai:04}.
Whereas other Fermi surface techniques rely on a long
mean-free-path of the electron or relatively defect-free crystal,
Compton scattering is a robust technique insensitive to defects or disorder,
providing a one-dimensional projection (double integral)
of the underlying bulk electron momentum distribution.
For each Compton profile $\sim300~000$ counts
in the peak data channel were accumulated. Of the twenty-four Compton profiles
that were collected, three were of high-symmetry directions, namely [100], [110]
and [111], and ten were the `special directions' outlined by Kontrym-Sznajd
{\it et al.} \cite{kontrymsznajd:02}; the remaining eleven Compton profiles were
chosen in such a way so as to be equally spaced throughout the Brillouin zone.
Each Compton profile was corrected for
possible multiple-scattering contributions using a Monte Carlo method
\cite{chomilier:85}.

\begin{figure}[t]
\includegraphics[width=0.80\linewidth,clip]{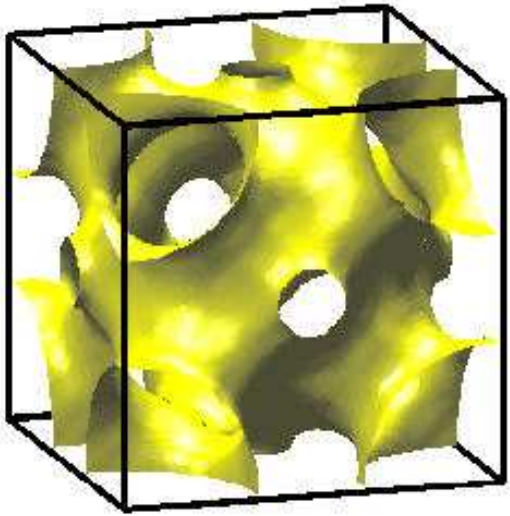}
\caption{\label{exp} (color online) The experimental Fermi surface of
Ni$_{0.68}$Al$_{0.32}$, determined from the momentum density reconstruction of
24 Compton profiles along different crystallographic directions. The symmetry
points are the same as those shown in Fig.\ \ref{kkr}.}
\end{figure}

\begin{figure}[t]
\includegraphics[width=0.80\linewidth,clip]{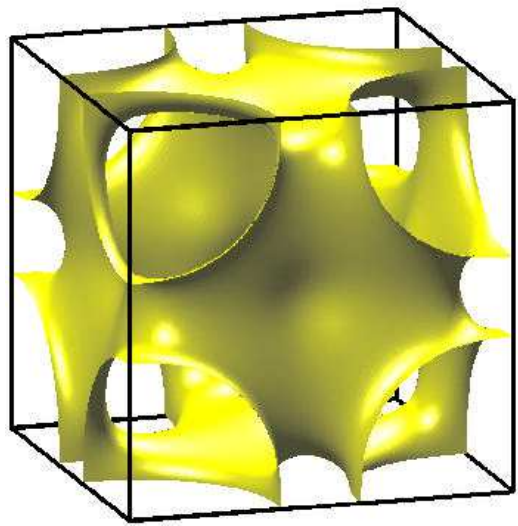}
\caption{\label{the_recon} (color online) The Fermi surface of
Ni$_{0.68}$Al$_{0.32}$, determined from the momentum density reconstruction of
24 Compton profiles along different crystallographic directions calculated
using the KKR method. The symmetry points are the same as those shown
in Fig.\ \ref{kkr}.}
\end{figure}

A three-dimensional momentum
density was reconstructed \cite{kontrymsznajd:00} from this set
of 24 profiles and then folded back into the first Brillouin zone using
the Lock-Crisp-West procedure \cite{lock:73} to obtain
the occupation density.
The experimental Fermi surface (shown in Fig.\ \ref{exp}) was extracted by
contouring this density at a level fixed by an extremum in the first derivative
along a direction where our calculations indicated the
Fermi surface was likely to be well-defined \cite{major:04}.
To illustrate that the reconstruction procedure does not introduce artefacts,
the Fermi surface reconstructed from the Compton profiles calculated
by the KKR method along the same 24 directions is shown in 
Fig.\ \ref{the_recon}.

\begin{figure}[b]
\includegraphics[width=0.80\linewidth,clip]{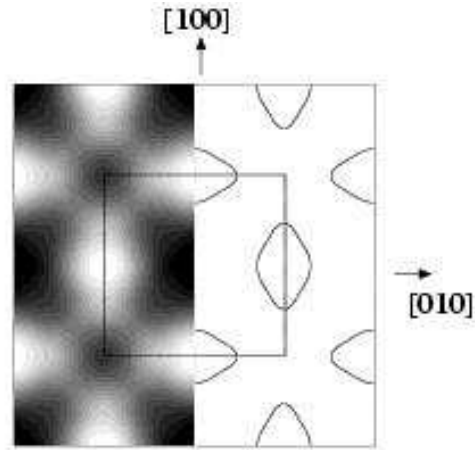}
\caption{\label{expslice} A slice through the
$k_{z}=0.48~(\pi/a)$ plane of the experimental data, for comparison
with Fig.\ \ref{kkr}(b). Shown on the
left is the occupation density through this slice, where brighter
shades represent a larger occupation. On the right is a contour of
the occupation density at the level corresponding to the Fermi energy.
The edge of the first Brillouin zone is marked.}
\end{figure}

Before addressing the issue of how faithfully the {\it ab initio} calculations
have described the observed Fermi surface, it is important to investigate
the nesting properties of the experimentally determined one. A plane-by-plane
inspection of the Fermi surface throughout the Brillouin zone revealed
that a vector
$\approx 0.18~[1,1,0](2\pi/a)$ connects a large area in the
manner predicted by Zhao and Harmon \cite{zhao:92}. The plane
through the BZ at $k_{z}=0.48~(\pi/a)$, for comparison with
Fig.\ \ref{kkr}(b), is shown in Fig.~\ref{expslice}.

We now return to the topology of the experimental Fermi surface, and to
the question of whether the {\it ab initio} calculations provide a good
description of the Fermi surface for these materials.
Clearly, the general shape of the experimental Fermi surface
(Fig.\ \ref{exp}) agrees well with the calculation (Fig.\ \ref{kkr}).
As discussed above, the regions of Fermi surface
responsible for the nesting are observed experimentally to be
relatively flat. This, in conjunction with the large density of states
(predicted by the calculations) spanning these wavevectors,
provides a large propensity
for nesting. There is, however, some noteworthy discrepancy between the 
calculated and experimental Fermi surfaces. Experimentally, a neck is
observed to open around the $X$-point of the Brillouin zone whereas according
to the calculations this sheet remains closed.
The calculated bandstructure reveals flat (almost dispersionless) bands along 
$X - M$ and $M - R$, lying just below the Fermi level, leading to a van Hove
singularity in the density of states at that energy. The opening up
of a Fermi surface neck along $\Gamma - X$ implies that the Fermi level
has crossed below this van Hove singularity and may be indicative of the
impending lattice instability at the martensitic transformation (where the
Fermi surface would undergo more substantial rearrangement).

In conclusion, we have presented the experimental Fermi surface of the
disordered alloy Ni$_{0.62}$Al$_{0.38}$ from the results of Compton
scattering experiments, providing evidence in support of the
intimate link between the electronic structure and the observed
phonon softening. The Fermi surface obtained from {\it ab initio} calculations
within the KKR-CPA framework has been shown to be in good agreement with
the experimentally determined Fermi surface, with the exception of the
opening of a neck along the $X$ direction of the Brillouin zone. It is
tentatively suggested that this topological disagreement may be explained
by the proximity of the Fermi level to a van Hove singularity in the density of
states.

\section*{Acknowledgements}
We acknowledge the financial support of the Royal Society (S.B.D. and M.S-C.),
the UK EPSRC, the Polish Academy of Sciences, the Japan Society for the
Promotion of Science (R.J.W.) and CELTAM in Poland (M.S-C.).
This experiment was performed with the approval of the Japan
Synchrotron Radiation Research Institute (JASRI)
(Proposal No. 2004A0332-ND3a-np). We are indebted to the LACMS (Bristol)
for access to their Beowulf cluster.
We would also like to thank H.~Ebert for useful discussions, and Akiko Kikkawa
(RIKEN/SPring-8) for characterisation of the sample.

\end{document}